\documentclass[pre,twocolumn,aps]{revtex4}
\newcommand{\bec}[1]{\mbox{\boldmath $ #1$}}
\usepackage{graphicx}
\begin{document}
\bigskip
\bigskip
\title{Nonhelical mean-field dynamos in a sheared turbulence}
\author{Igor Rogachevskii}
\email{gary@bgu.ac.il} \homepage{http://www.bgu.ac.il/~gary}
\author{Nathan Kleeorin}
\email{nat@bgu.ac.il}
\affiliation{Department of Mechanical Engineering, The Ben-Gurion University of the Negev,
 POB 653, Beer-Sheva 84105, Israel}
\date{\today}
\begin{abstract}
Mechanisms of nonhelical large-scale dynamos (shear-current dynamo and effect of homogeneous kinetic helicity fluctuations with zero mean) in a homogeneous turbulence with large-scale shear are discussed. We have found that the shear-current dynamo can act even in random flows with small Reynolds numbers. However, in this case mean-field dynamo requires small magnetic Prandtl numbers (i.e., ${\rm Pm} < {\rm Pm}^{\rm cr}<1$). The threshold in the magnetic Prandtl number, ${\rm Pm}^{\rm cr} = 0.24$, is determined using  second order correlation approximation (or first-order smoothing approximation) for a background random flow with a scale-dependent viscous correlation time $\tau_c=(\nu k^2)^{-1}$ (where $\nu$ is the kinematic viscosity of the fluid and $k$ is the wave number). For turbulent flows with large Reynolds numbers shear-current dynamo occurs for arbitrary magnetic Prandtl numbers. This dynamo effect represents a very generic mechanism for generating large-scale magnetic fields in a broad class of astrophysical turbulent systems with large-scale shear. On the other hand, mean-field dynamo due to homogeneous kinetic helicity fluctuations alone in a sheared turbulence is not realistic for a broad class of astrophysical systems because it requires a very specific random forcing of kinetic helicity fluctuations that contains, e.g., low-frequency oscillations.

\end{abstract}

\maketitle

\section{Introduction}

An origin of solar, stellar and galactic large-scale magnetic fields
is related to a mean-field dynamo. This dynamo can be driven by the joint action of small-scale turbulent flows with a nonzero mean kinetic helicity and large-scale differential rotation (see, e.g., Moffatt 1978; Parker 1979; Krause and R\"{a}dler 1980; Zeldovich et al. 1983; Ruzmaikin et al. 1988;  Ossendrijver 2003; R\"{u}diger and Hollerbach 2004; Brandenburg and Subramanian 2005).

Recent numerical experiments by Yousef et al. (2008) have demonstrated existence of a nonhelical large-scale dynamo in a turbulence with superimposed linear shear in elongated shearing boxes whereby mean $\alpha$ effect vanishes. The exponential growth of magnetic field has been found at scales which are much larger than the outer scale of the turbulence. An earlier indications of nonhelical turbulence amplifying large-scale magnetic field in the presence of a large-scale shear associated with mean flows has been found by Brandenburg (2005) and Brandenburg et al. (2005) in numerical experiments that used constant-in-time sinusoidal forcing functions. This implies that the
amplification effect in a sheared nonhelical turbulence appears to be numerically robust. Note also that numerical experiments with Taylor-Green forcing is another example of a mean-field dynamo produced by a combined effect of a nonhelical turbulence and a complicated large-scale flow (Ponty et al. 2005).

One of the possible mechanism of the nonhelical large-scale dynamo
in a homogeneous sheared turbulence is a shear-current dynamo effect (see Rogachevskii and Kleeorin 2003; 2004; 2007; Rogachevskii et al. 2006a, 2006b). The physics of this phenomenon is following. Upward and downward turbulent eddies result in deformations of the original nonuniform magnetic field lines. In a turbulence with a large-scale shear the inhomogeneity of the original mean magnetic field
breaks a symmetry between the influence of the upward and downward
turbulent eddies on the mean magnetic field. This causes the mean
electric current along the mean magnetic field and results in  systematic amplification of the large-scale magnetic field.

The shear-current dynamo has been  previously studied for large Reynolds numbers using the spectral tau-approximation (see Rogachevskii and Kleeorin 2003; 2004; 2007). On the other hand, in a random flow with small Reynolds numbers the dynamo action in nonhelical shear flows has not yet been found in the framework of the second order correlation approximation or first-order smoothing approximation (see R\"{a}dler and Stepanov 2006; R\"{u}diger and Kichatinov 2006). How generic is the latter result? In particular, how the dynamo action may depend on the spectral properties of a random flow with small Reynolds numbers? One of the goals of this study is to revise this problem for the case of a random flow with small Reynolds numbers.

Another effect that might explain the large-scale dynamo in a sheared turbulence with a zero mean kinetic helicity is associated with kinetic helicity fluctuations. Dynamics of large-scale magnetic field in the presence of kinetic helicity fluctuations with a zero mean in a shear-free turbulence has been studied for the first time by Kraichnan (1976). This problem is formulated in the following way. Let us consider a small-scale turbulence produced by a  random forcing ${\bf F}^{(u)}$ located in small scales $l_\nu \ll l_{\rm turb} \ll l_0$ (and $\tau_\nu \ll \tau_{\rm turb} \ll \tau_0)$, while in larger scales $l_0 \ll l \ll l_\chi$ (and $\tau_0 \ll \tau \ll \tau_\chi$) there are kinetic helicity fluctuations (or $\tilde \alpha$ fluctuations) with a zero mean produced by a  random forcing ${\bf F}^{(\chi)}$.

The mean-field effects occur at very large scales $L \gg l_\chi$ (and times $\tau_{_{L}} \gg \tau_\chi$), where the mean kinetic helicity is zero. All mean quantities are determined by double averaging over velocity fluctuations, $\langle ... \rangle$, and over kinetic helicity fluctuations $\langle ... \rangle^{(\alpha)}$ [see detailed discussion by Sokolov (1997) about various mathematical aspects of this problem]. Numerical simulations of the magnetic field evolution in accretion discs by Vishniac and Brandenburg (1997) have demonstrated that kinetic helicity fluctuations with a zero mean can result in generation of large-scale magnetic field (see also Brandenburg et al. 2008).

Let us discuss theoretical aspects of this problem. In a shear-free turbulence kinetic helicity fluctuations cause two effects: (i) a negative contribution to the turbulent magnetic diffusion, $\eta^{(\alpha)}_{_{T}} = - \tau_\chi \, \langle \tilde{\alpha}^2 \rangle^{(\alpha)}$; and (ii) a large-scale drift velocity of the mean magnetic field, ${\bf V}^{(\alpha)} \propto \tau_\chi \, \bec{\nabla} \langle \tilde{\alpha}^2 \rangle^{(\alpha)}$ (see Kraichnan 1976; Moffatt 1978). In a turbulence with large-scale shear, inhomogeneous kinetic helicity fluctuations can produce a mean-field dynamo (Silant'ev 2000). Indeed, a combined effect of the inhomogeneous fluctuations and large-scale shear superimposed on turbulence, produces a nonzero mean alpha effect: $\bar{\alpha}^{(S,\alpha)} \propto - \tau_\chi^2 \, S \, {\nabla}_z \langle \tilde{\alpha}^2 \rangle^{(\alpha)}$, while $\langle \tilde{\alpha} \rangle^{(\alpha)}=0$. Here the mean vorticity due to the large-scale shear is $\bar {\bf W} = S \, {\bf e}_z$. The equation for $\bar{\alpha}^{(S,\alpha)}$ has been derived using the second order correlation approximation and the spectral tau-approximation (see for details, Kleeorin and Rogachevskii 2008). The large-scale shear and the mean alpha effect can result in the mean-field dynamo that acts similarly to the $\alpha\Omega$-dynamo.

Using phenomenological arguments, Proctor (2007) has suggested that  homogeneous kinetic helicity fluctuations in a homogeneous turbulence with a large-scale shear may generate a large-scale magnetic field. Such possibility for a large-scale dynamo has been recently examined by Kleeorin and Rogachevskii (2008) using the second order correlation approximation and the spectral tau-approximation. This study has not found large-scale dynamo produced by homogeneous kinetic helicity fluctuations alone with zero mean value in a sheared homogeneous turbulence. However, how generic is the latter statement? One of the goals of this study is to revise this problem. We have demonstrated that only for a specific random forcing of kinetic helicity fluctuations that also contains low-frequency oscillations, the large-scale dynamo in a homogeneous turbulence with a large-scale shear might be possible.

This paper is organized as follows. In Sec.~II we investigate shear-current dynamo for a random flow with small Reynolds numbers and different spectral properties. In Sec.~III we study the effect of homogeneous kinetic helicity fluctuations with a zero mean in a sheared turbulence. In Sec.~IV we draw concluding remarks.

\section{The shear-current effect}

In order to study the shear-current effect in a random flow with small Reynolds numbers we use a second order correlation approximation (SOCA). This approximation is valid only for small hydrodynamic Reynolds numbers. Even in a highly conductivity limit (large magnetic Reynolds numbers), SOCA can be valid only for small Strouhal numbers (i.e., for very short correlation time).

We use equation of motion and induction equation for fluctuations of velocity and magnetic fields, exclude the pressure term from the equation of motion by calculation $\bec{\nabla} {\bf \times} (\bec{\nabla} {\bf \times} {\bf u})$, where ${\bf u}$ are velocity fluctuations. We rewrite the obtained equation and the induction equation in a Fourier space and apply the two-scale approach (i.e., we use large-scale and small-scale variables). We neglect nonlinear terms but keep molecular dissipative terms in the equations for fluctuations of velocity and magnetic fields. We seek for a solution of the obtained equations for fluctuations of velocity, ${\bf u}$, and magnetic, ${\bf b}$, fields as an expansion for a weak velocity shear:
\begin{eqnarray}
{\bf u}({\bf k}, \omega) &=& {\bf u}^{(0)}({\bf k}, \omega) + {\bf u}^{(1)}({\bf k}, \omega) + ... \;,
\label{C4} \\
{\bf b}({\bf k}, \omega) &=& {\bf b}^{(0)}({\bf k}, \omega) + {\bf b}^{(1)}({\bf k}, \omega) + ... \;,
\label{C5}
\end{eqnarray}
where
\begin{eqnarray}
b_i^{(0)}({\bf k}, \omega) &=& G_\eta(k, \omega) \, \biggl[i({\bf k} {\bf \cdot} \bar {\bf B}) u_i^{(0)} - \Big(k_{m} \, {\partial u_i^{(0)} \over \partial k_{n}}
\nonumber\\
&& + \delta_{im} \, u_n^{(0)} \Big) (\nabla_{n} \bar B_{m}) \biggr] \;,
 \label{C1}\\
u_i^{(1)}({\bf k}, \omega) &=& G_\nu(k, \omega) \, \biggl[
2 k_{iq} \, u_p^{(0)} +  k_{q} {\partial u_i^{(0)}\over \partial k_{p}}
\nonumber\\
&& - \delta_{iq} \, u_p^{(0)} \biggr] (\nabla_{p} \bar U_{q}) \;,
\label{C2}\\
b_i^{(1)}({\bf k}, \omega) &=& G_\eta(k, \omega) \, \biggl\{ \biggl[i({\bf k} {\bf \cdot} \bar {\bf B}) \, u_i^{(1)} - \Big(k_{m} {\partial u_i^{(1)} \over \partial k_{n}}
\nonumber\\
&& + \delta_{im} \, u_n^{(1)} \Big) \, (\nabla_{n} \bar B_{m}) \biggr] + \biggl[k_{q} {\partial \, b_i^{(0)} \over \partial k_{p}}
\nonumber\\
&&  + \delta_{iq} \, b_p^{(0)} \biggr] (\nabla_{p} \bar U_{q}) \biggr\}   \; .
\label{C3}
\end{eqnarray}
Here $\bar{\bf U}$ and $\bar{\bf B}$ are the mean velocity and magnetic fields, ${\bf k}$ and $\omega$ are the wave number and frequency, $G_\nu(k, \omega) = (\nu k^2 - i \omega)^{-1}$ and $G_\eta(k, \omega) = (\eta k^2 - i \omega)^{-1}$, $\, \eta$ is the magnetic diffusion coefficient due to electrical conductivity of the fluid, and $\nu$ is the kinematic viscosity of the fluid. For derivation of Eqs.~(\ref{C1})-(\ref{C3}) we use an identity
\begin{eqnarray*}
\int \bar U_q({\bf Q}) \, b_n({\bf k} - {\bf Q}) \,d{\bf Q} = i (\nabla_{p} \bar U_{q}) \, {\partial b_n \over \partial k_{p}} \;,
\end{eqnarray*}
that is valid at least for a linear velocity field. Equations~(\ref{C1})-(\ref{C3}) coincide with that derived by R\"{a}dler and  Stepanov (2006). These equations allow us to determine the cross-helicity tensor $g_{mn}^{(1)} = \langle u_m^{(0)} \, b_n^{(1)}  \rangle + \langle u_m^{(1)} \, b_n^{(0)} \rangle$ and the contributions, ${\cal E}_{i}^{(S)} = \varepsilon_{imn} \, \int \, g_{mn}^{(1)}({\bf k}, \omega) \,d {\bf k} \,d \omega$, to the electromotive force caused by sheared turbulence.
For the integration in ${\bf k}$-space we have to specify a model for the background shear-free turbulence (with $\bar {\bf B} = 0)$, which is determined by equation:
\begin{eqnarray}
\langle u_i \, u_j \rangle^{(0)}({\bf k},\omega) = \langle {\bf u}^2 \rangle^{(0)} \, {P_{ij}(k) \, E(k) \over 8 \pi^2 \, k^{2} \, \tau_c \, (\omega^2 + \tau_c^{-2})} \;,
 \label{AB1}
\end{eqnarray}
where $E(k)$ is the energy spectrum (e.g., a power-law spectrum), $\tau_c$ is the correlation time, $P_{ij}(k) = \delta_{ij} - k_i k_j / k^2$ and $\delta_{ij}$ is the Kronecker tensor. This model corresponds to the correlation function: $\langle u_i(t) u_j(t+\tau) \rangle \propto \exp (-\tau/\tau_c)$. Straightforward calculations yields the contributions to the electromotive force caused by sheared turbulence:
\begin{eqnarray}
{\cal E}_{i}^{(S)} &=& l_0^2 \, [A_1 \, \varepsilon_{ipk} \, (\partial \bar U)_{pq} \, (\partial \bar B)_{qk} + A_2 \, \bar W_k \, (\partial \bar B)_{ik}
\nonumber\\
&&  + A_3 \, \bar J_k \, (\partial \bar U)_{ik} + A_4 \, (\bar {\bf W} {\bf \times} \bar {\bf J})_i] \;,
 \label{AB2}
\end{eqnarray}
where $(\partial \bar U)_{ij} = (\nabla_i \bar U_{j} + \nabla_j \bar U_{i}) / 2$, $\, \bar{\bf W}= \bec{\nabla} {\bf \times} \bar{\bf U}$ is the mean vorticity, $l_0$ is the maximum scale of turbulent motions (the energy containing scale), $\, \bar{\bf J}= \bec{\nabla} {\bf \times} \bar{\bf B}$ is the mean electric current, and the coefficients $A_n$  are given in Appendix. The equation for the evolution of the mean magnetic field, $\bar{\bf B} = (\bar B_x(z), \bar B_y(z), 0)$, reads
\begin{eqnarray}
{\partial \bar B_x \over \partial t} &=& - \sigma_{_{B}} \, S \, l_0^2 \, \bar B''_y + (\eta + \eta_{_{T}}) \, \bar B''_x  \;,
 \label{AB3}\\
{\partial \bar B_y \over \partial t} &=& S \, \bar B_x + (\eta + \eta_{_{T}}) \, \bar B''_y
\;,
 \label{AB4}
\end{eqnarray}
where we use linear velocity shear $\bar {\bf U} = (0, Sx, 0)$, $\,\bar B''_i = \partial^2 \bar B_i / \partial z^2 $, $\, \eta_{_{T}} \propto \tau_0 \, \langle {\bf u}^2 \rangle^{(0)}$ is the turbulent magnetic diffusion coefficient, $\tau_0 = l_0 / \sqrt{\langle {\bf u}^2 \rangle^{(0)}}$, and we neglect small contributions to the coefficient of turbulent magnetic diffusion caused by the shear motions because we consider a small shear, $S \tau_0 \ll 1$. The coefficient $\sigma_{_{B}}$ entering in Eq.~(\ref{AB3}) is given by
\begin{eqnarray}
\sigma_{_{B}} &=& {\nu \over 15 \pi \tau_0^2} \, \int  [I_4 - \nu k^2  I_3 + \eta k^2 (I_1
- I_2)] \, E(k)  \, k^{2} \,dk \;,
\nonumber\\
 \label{AB5}
\end{eqnarray}
the functions $I_n(k)$ for $\tau_c=1 / (\nu k^2)$ are given in Appendix. Using the explicit form of the functions $I_n(k)$, we obtain the following expression for the coefficient $\sigma_{_{B}}$:
\begin{eqnarray}
\sigma_{_{B}} &=& {1 \over 60 \, (\tau_0 \,\nu)^2} \, {\rm Pm} \, (1 - 4 {\rm Pm} - {\rm Pm}^2) \, \int {E(k) \over k^{4}}  \,dk ,
\nonumber\\
 \label{AB6}
\end{eqnarray}
where ${\rm Pm} = \nu/\eta$ is the magnetic Prandtl number.
The solution of Eqs.~(\ref{AB3}) and~(\ref{AB4}) we seek for in
the form $ \propto \exp(\gamma_{_{B}} \, t + i K_z \, z) ,$ where the growth rate, $\gamma_{_{B}}$, of the mean magnetic field is given by
\begin{eqnarray}
\gamma_{_{B}} = S \, l_0 \, \sqrt{\sigma_{_{B}}} \, K_z -
(\eta + \eta_{_{T}}) \, K_z^2 \; .
 \label{AB7}
\end{eqnarray}
and $\sigma_{_{B}}>0$ when ${\rm Pm} < 0.24$.

In the present study we use the SOCA procedure that is valid only for ${\rm Re} \ll 1$. It follows from Eqs.~(\ref{AB6}) and~(\ref{AB7}) that for ${\rm Re} \ll 1$, the dynamo instability due to the shear-current effect occurs when ${\rm Pm} < 0.24$ (i.e., for small magnetic Prandtl numbers). This result has been obtained for a model of the background shear-free turbulence determined by Eq.~(\ref{AB1}) with $\tau_c=1 / (\nu k^2)$.
Note that R\"{a}dler and  Stepanov (2006) used a model of the background shear-free turbulence with a constant scale-independent correlation time $\tau_c$. A possibility for the shear-current dynamo for small magnetic Prandtl numbers in the case of ${\rm Re} \ll 1$ has been pointed out by R\"{u}diger (2007), although this was not explicitly mentioned in his previous study using the SOCA procedure and a more simple model for the background shear-free turbulence:
$\langle u_i \, u_j \rangle^{(0)}({\bf k},\omega) \propto \langle {\bf u}^2 \rangle^{(0)} \, P_{ij}(k) \, E(k) \, \delta(\omega)$
(see R\"{u}diger and Kitchatinov 2006). For turbulent flows with large Reynolds numbers shear-current dynamo occurs for arbitrary magnetic Prandtl numbers (see Rogachevskii and Kleeorin 2003; 2004; 2007).

\section{Effect of kinetic helicity fluctuations}

In order to study effect of kinetic helicity fluctuations with a zero mean on large-scale dynamo we use a second order correlation approximation. This procedure yields the equation for the evolution of the magnetic field ${\bf B}$:
\begin{eqnarray}
{\partial {\bf B} \over \partial t} = \bec{\nabla} {\bf \times} \Big(
\tilde\alpha {\bf B} + {\bf V} {\bf \times} {\bf B} - (\eta + \eta_{_{T}}) {\bf J}\Big) + {\bf B}^N \;,
\label{B5}
\end{eqnarray}
where ${\bf J}= \bec{\nabla} {\bf \times} {\bf B}$ is the electric current, ${\bf B}^N$ are the nonlinear terms, ${\bf V}+{\bf u}$ is the total velocity and $\langle {\bf u} \rangle=0$. In this section we do not consider the shear-current effect.

In the scales $l_0 \ll l \ll l_\chi$ there are fluctuations of $\tilde \alpha$. Let us consider homogeneous kinetic helicity fluctuations. In order to derive equation for the the mean magnetic field $\bar{\bf B}=\langle {\bf B} \rangle^{(\alpha)}$, we determine the contribution to the mean electromotive force caused by the sheared turbulence and the kinetic helicity fluctuations, ${\cal E}_{j}^{(S,\alpha)} = \langle \tilde\alpha B_j \rangle^{(\alpha)}$. To this end we use Eq.~(\ref{B5}) in which we neglect the nonlinear terms ${\bf B}^N$. Solving this equation in a Fourier space we determine the magnetic field $B_j({\bf k}, \omega)$, where the wave vector ${\bf k}$ and the frequency $\omega$ are located in the spatial scales $l_0 \ll l \ll l_\chi$ and in the time scales $\tau_0 \ll \tau \ll \tau_\chi$. Multiplying the magnetic field $B_y({\bf k}, \omega)$ by $\tilde \alpha$ and averaging over kinetic helicity fluctuations we determine ${\cal E}_{y}^{(S,\alpha)}$:
\begin{eqnarray}
{\cal E}_{y}^{(S,\alpha)} = S \, \bar J_x \, \int G_T^2(k,\omega) \, f_{\alpha}(k,\omega) \, dk \, d\omega  \;,
 \label{C14}
\end{eqnarray}
where $G_T(k,\omega) = [(\eta + \eta_{_{T}}) k^2 - i \omega]^{-1}$, $\, f_{\alpha}(k,\omega) = \langle \tilde \alpha(\omega) \tilde \alpha(-\omega) \rangle^{(\alpha)}$, $\, \bar {\bf U} = (0, Sx, 0)$ is the background shear velocity, and $\bar{\bf J}= \bec{\nabla} {\bf \times} \bar{\bf B}$ is the mean electric current. We assume that the mean magnetic field has the form: $\bar{\bf B} = (\bar B_x(z), \bar B_y(z), 0)$,  and neglect small contributions $\sim O(\tau_0 /\tau_\chi)$ to the mean electromotive force ${\cal E}_{y}^{(S,\alpha)}$.

We use the following model for the spectral function $f_{\alpha}(k,\omega)$:
\begin{eqnarray}
f_{\alpha}(k,\omega) = \langle \tilde \alpha^2 \rangle^{(\alpha)} \, {E_{\alpha}(k) \over \pi \tau_\chi (\omega^2 + \tau_\chi^{-2})} \; .
 \label{AC1}
\end{eqnarray}
This model corresponds to the following correlation function $\langle \tilde \alpha(t) \tilde \alpha(t+\tau) \rangle^{(\alpha)} \propto \exp (-\tau/\tau_\chi)$. In earlier studies by Kleeorin and Rogachevskii (2008), a more simple model for the spectral function has been used:
$f_{\alpha}(k,\omega) = \langle \tilde \alpha^2 \rangle^{(\alpha)} \, E_{\alpha}(k) \delta(\omega)$.

The contribution to the mean electromotive force caused by the sheared turbulence and the kinetic helicity fluctuations is given by ${\cal E}_{j}^{(S,\alpha)} = - \sigma_\alpha \, S \, \langle \tilde \alpha^2 \rangle^{(\alpha)} \, \tau_\chi^2 \, B'_y$, where the parameter
\begin{eqnarray}
\sigma_\alpha = \int {E_{\alpha}(k) \over [1 + \tau_\chi \, (\eta + \eta_{_{T}}) \, k^2]^2} \,dk > 0 \; .
 \label{AC2}
\end{eqnarray}
Here we use an identity $\int G^2_\eta \,G_a \,G^*_a \, d\omega = \pi / [a \, (\eta \, k^2 + a)^2]$, where $G_a(k,\omega) = (a - i \omega)^{-1}$ with $a=\tau_\chi^{-1}$.
The equation for the evolution of the mean magnetic field, $\bar{\bf B} = (\bar B_x(z), \bar B_y(z), 0)$, reads
\begin{eqnarray}
{\partial \bar B_x \over \partial t} &=& \sigma_\alpha \, S \, \langle \tilde \alpha^2 \rangle^{(\alpha)} \, \tau_\chi^2 \, \bar B''_y + \tilde \eta_{_{T}} \, \bar B''_x  \;,
 \label{E2}\\
{\partial \bar B_y \over \partial t} &=& S \, \bar B_x + \tilde \eta_{_{T}} \, \bar B''_y
\;,
 \label{E3}
\end{eqnarray}
where $\tilde \eta_{_{T}} = \eta + \eta_{_{T}} + \eta^{(\alpha)}_{_{T}}$. Here we neglect small contributions to the coefficient of turbulent magnetic diffusion caused by the shear motions because $S \tau_0 \ll 1$. Note that for enough general model~(\ref{AC1}) of the spectral function $f_{\alpha}(\omega, k)$, the parameter $\sigma_\alpha$ is always positive. It is also positive when $\tau_\chi=(\nu k^2)^{-1}$ (see the spectral model used in Sect. 2). This implies that homogeneous kinetic helicity fluctuations alone with zero mean value for general model~(\ref{AC1}) in a sheared homogeneous turbulence cannot cause a large-scale dynamo [see Eqs.~(\ref{E2}) and~(\ref{E3})].

However, for a specific random forcing of kinetic helicity fluctuations that also contains low-frequency oscillations, e.g.,
\begin{eqnarray}
\langle \tilde \alpha(t) \tilde \alpha(t+\tau) \rangle^{(\alpha)} \propto \exp (-\tau/\tau_\chi) \, \cos(\omega_{\rm w} \tau) \;,
 \label{E4}
\end{eqnarray}
there is a possibility for a large-scale dynamo action due to homogeneous kinetic helicity fluctuations in a sheared homogeneous turbulence. In this case the spectral function $f_{\alpha}(k,\omega)$ is given by:
\begin{eqnarray*}
f_{\alpha}(k,\omega) = \langle \tilde \alpha^2 \rangle^{(\alpha)} \, {E_{\alpha}(k) \over 2\pi} \, \biggl[{a \over \omega^2 + a^2} + {a^* \over \omega^2 + (a^*)^2} \biggr] ,
\end{eqnarray*}
with $a=\tau_\chi^{-1} + i\omega_{\rm w}$, and the parameter $\sigma_\alpha$,
\begin{eqnarray}
\sigma_\alpha &=&  \int  {[1 + \tau_\chi \, (\eta + \eta_{_{T}}) \, k^2]^2 - (\omega_{\rm w} \tau_\chi)^2 \over \{[1 + \tau_\chi \, (\eta + \eta_{_{T}}) \, k^2]^2 + (\omega_{\rm w} \tau_\chi)^2\}^2}
\, E_{\alpha}(k) \,dk \;,
\nonumber\\
\label{AC3}
\end{eqnarray}
is negative when $\omega_{\rm w} \tau_\chi > 1 + \tau_\chi \, (\eta + \eta_{_{T}}) \, k^2$. This implies that for model~(\ref{E4}) a large-scale dynamo due to homogeneous kinetic helicity fluctuations in a sheared homogeneous turbulence can occurs. However, this model for the function $f_{\alpha}(k,\omega)$ of kinetic helicity fluctuations seems to be not realistic.

\section{Conclusions}

Two types of nonhelical large-scale dynamos due to shear-current effect and homogeneous kinetic helicity fluctuations with zero mean in a sheared turbulence are investigated using a second order correlation approximation. The mechanism for the shear-current dynamo is following. The large-scale velocity shear creates anisotropy of turbulence that produces a contribution to the electromotive force, $\bar {\bf W} {\bf \times} \bar {\bf J}$, caused by the shear. Joint effects of the electromotive force $\bar {\bf W} {\bf \times} \bar {\bf J}$ and stretching of the mean magnetic field due to the large-scale shear motions cause the shear-current dynamo instability.
This effect occurs even for small Reynolds numbers. However, the dynamo instability in this case requires small magnetic Prandtl numbers (${\rm Pm} < 0.24$). This dynamo threshold is found for a model of a  random flow with the correlation time $\tau_c=(\nu k^2)^{-1}$. The shear-current dynamo for large Reynolds numbers is independent of magnetic Prandtl numbers.

Another possible mechanism for the nonhelical large-scale dynamo is associated with homogeneous kinetic helicity fluctuations in a sheared turbulence. However, this kind of mean-field dynamo is not universal and can occur only for a specific random forcing of kinetic helicity fluctuations that contains, e.g., low-frequency oscillations.

The discussed effects in this study might be important in a broad class of astrophysical flows. For instance, sheared turbulence is a universal feature in astrophysical flows, e.g., in stellar interiors, accretion disks, irregular galaxies \, \, (Balbus and Hawley 1998;  Chyzy et al. 2000;  Ossendrijver 2003; Brandenburg and Subramanian 2005; Donati et al. 2005; Gaensler et al. 2005), and in liquid-metal laboratory dynamo experiments (see, e.g., Monchaux et al. 2007).

Non-symmetrical explosions of supernova may produce fluctuations of kinetic helicity located in larger scales than  small-scale turbulence existing in convective zones inside stars. On the other hand, the shear-current dynamo acts together with the $\alpha$-shear dynamo. The shear-current effect does not quenched (see Rogachevskii and Kleeorin 2004; Rogachevskii et al. 2006b) contrary to the quenching of the nonlinear $\alpha$ effect, the turbulent magnetic diffusion, etc. This implies that the shear-current dynamo might be the only surviving effect, which can explain the origin of large-scale magnetic fields in sheared astrophysical turbulence.

\acknowledgements
We have benefited from stimulating discussions with E.~Blackman, A.~Brandenburg, K.-H.~R\"{a}dler, G.~R\"{u}diger, A.~Schekochihin and K.~Subramanian. This work has benefited from Nordita program ''Turbulence and Dynamos'' and KITP program ''Dynamo Theory''. This research was supported in part by the National Science Foundation under Grant No. PHY05-51164.

\appendix

\section{Coefficients $A_n$ and functions $I_n(k)$}

The coefficients $A_n$ entering in Eq.~(\ref{AB2}) are given by
\begin{eqnarray*}
A_1 &=& {\nu \over 30 \, \pi} \, \int \, [2 I_5 - 2 I_4 + 5 I_8 + 2 \nu k^2 (2 I_3 + I_6)
\\
&&  + 2 \eta k^2 (2 I_2 - I_7)] \, E(k)  \, k^{2}  \,dk \;,
\\
A_2 &=& {\nu \over 60\, \pi} \, \int \, [6 I_4 + 5 I_8 - 6 \nu k^2 I_3 + 2 \eta k^2 (3 I_1 - 3 I_2
\\
&& - 2I_7)] \, E(k) \, k^{2} \, dk\;,
\\
A_3 &=& {\nu \over 60\, \pi} \, \int \, [5 I_8 - 2 I_5 - 2 \nu k^2 (I_3 + I_6)
\\
&& - 2 \eta k^2 (I_1 + I_2 + I_7)] \, E(k) \, k^{2}  \,dk\;,
\\
A_4 &=& - {\nu \over 24\, \pi} \, \int \, I_8 \, E(k) \, k^{2}  \,dk \;,
\end{eqnarray*}
where the functions $I_n(k)$ for $\tau_c^{-1}=\nu k^2$ are given by
\begin{eqnarray*}
I_1(k) &=& \int G^2_\eta \,G^2_\nu \,G^*_\nu \, d\omega = {\pi \over 2 \, \nu^2 \, (\nu + \eta)^2 \, k^8} \;,
\\
I_2(k) &=& \int G^2_\eta \,G_\nu \,(G^*_\nu)^2 \, d\omega = {\pi \, (5 \nu + \eta) \over 2 \, \nu^2 \, (\nu + \eta)^3 \, k^8} \;,
\\
I_3(k) &=& \int G_\eta \,G_\nu \,(G^*_\nu)^3 \, d\omega = {\pi \over 4 \, \nu^3 \, (\nu + \eta)^3 \, k^8}
\\
&& \times \, [2 \nu (\nu + \eta)+ (\nu + \eta)^2 + 4 \nu^2] \;,
\\
I_4(k) &=& \int G_\eta \,G_\nu \,(G^*_\nu)^2 \, d\omega = {\pi \, (3 \nu + \eta) \over 2 \, \nu^2 \, (\nu + \eta)^2 \, k^6} \;,
\\
I_5(k) &=& \int G_\eta \,G^2_\nu \,G^*_\nu \, d\omega = {\pi \over 2 \, \nu^2 \, (\nu + \eta) \, k^6} \;,
\\
I_6(k) &=& \int G_\eta \,G^3_\nu \,G^*_\nu \, d\omega = {\pi \over 4 \, \nu^3 \, (\nu + \eta) \, k^8} \;,
\\
I_7(k) &=& \int G^3_\eta \,G_\nu \,G^*_\nu \, d\omega = {\pi \over \nu \, (\nu + \eta)^3 \, k^8} \;,
\\
I_8(k) &=& \int G^2_\eta \,G_\nu \,G^*_\nu \, d\omega = {\pi \over \nu \, (\nu + \eta)^2 \, k^6} \; .
\end{eqnarray*}

\end{document}